\documentclass[a4,12pt]{article}
\usepackage[dvips]{graphicx,psfrag}
\usepackage{float}
\def\ssW{\scriptscriptstyle W}
\def\sla#1{\mbox{$#1\hspace*{-0.17cm}\scriptstyle{/}\:$}}

\def\ls{\lower0.5ex\hbox{$\buildrel >\over{\scriptstyle\sim}$}}
\def\rs{\lower0.5ex\hbox{$\buildrel <\over{\scriptstyle\sim}$}} 
\def\non{\nonumber}
\begin{document}
\pagestyle{empty} \setlength{\footskip}{2.0cm}
\setlength{\oddsidemargin}{0.5cm}
\setlength{\evensidemargin}{0.5cm}
\renewcommand{\thepage}{-- \arabic{page} --}
\def\mib#1{\mbox{\boldmath $#1$}}
\def\bra#1{\langle #1 |}  \def\ket#1{|#1\rangle}
\def\vev#1{\langle #1\rangle} \def\dps{\displaystyle}
\newcommand{\fcal}{{\cal F}}
\newcommand{\gcal}{{\cal G}}
\newcommand{\ocal}{{\cal O}}
\newcommand{\El}{E_\ell}
\renewcommand{\thefootnote}{$\sharp$\arabic{footnote}}
\newcommand{\W}{{\scriptstyle W}}
 \newcommand{\I}{{\scriptscriptstyle I}}
 \newcommand{\J}{{\scriptscriptstyle J}}
 \newcommand{\K}{{\scriptscriptstyle K}}
%
 \def\thebibliography#1{\centerline{REFERENCES}
 \list{[\arabic{enumi}]}{\settowidth\labelwidth{[#1]}\leftmargin
 \labelwidth\advance\leftmargin\labelsep\usecounter{enumi}}
 \def\newblock{\hskip .11em plus .33em minus -.07em}\sloppy
 \clubpenalty4000\widowpenalty4000\sfcode`\.=1000\relax}\let
 \endthebibliography=\endlist
 \def\sec#1{\addtocounter{section}{1}\section*{\hspace*{-0.72cm}
 \normalsize\bf\arabic{section}.$\;$#1}\vspace*{-0.3cm}}
\def\secnon#1{\section*{\hspace*{-0.72cm}
 \normalsize\bf$\;$#1}\vspace*{-0.3cm}}
 \def\subsec#1{\addtocounter{subsection}{1}\subsection*{\hspace*{-0.4cm}
 \normalsize\bf\arabic{section}.\arabic{subsection}.$\;$#1}\vspace*{-0.3cm}}
\vspace*{-1.7cm}
\begin{flushright}
$\vcenter{
\hbox{{\footnotesize TOKUSHIMA Report}}
{ \hbox{(arXiv:1508.01250)}  }
}$
\end{flushright}

\vskip 1.3cm
\begin{center}
  {\large\bf Decoupling theorem in top productions/decays revisited}

\vskip 0.26cm
  {\large\it -- To what extent can we understand it visually? --}

\end{center}

\vspace{0.5cm}
\begin{center}
\renewcommand{\thefootnote}{*)}
Zenr\=o HIOKI \footnote{E-mail address: \tt hioki@tokushima-u.ac.jp}
\end{center}

\vspace*{0.5cm}
\centerline{\sl Institute of Theoretical Physics,\ University of Tokushima}

\vskip 0.2cm
\centerline{\sl Tokushima 770-8502, Japan}

\vskip 0.2cm

\vspace*{3.4cm}
\centerline{ABSTRACT}

\vspace*{0.2cm}
\baselineskip=21pt 
I revisit here the decoupling theorem in top-quark productions/decays, which
states that the angular distribution of any final-particle produced in those
processes does not depend on any possible nonstandard top-quark decay
interactions at their leading order when certain conditions are
satisfied. Towards a simple, intuitive and visual understanding of this
theorem, I will study to what extent we could explain why such a theorem
holds without relying on any specific/detailed calculations.

\vspace*{0.5cm}


\vfill

\setcounter{page}{0}
\newpage
\renewcommand{\thefootnote}{$\sharp$\arabic{footnote}}
\pagestyle{plain} \setcounter{footnote}{0}

\sec{Introduction}

The top quark, the heaviest elementary particle of all those we have ever
encountered, has a mass very close to the electroweak scale, which makes us
expect that it plays an important role for understanding the spontaneous
breakdown of this symmetry and works as a window for new more fundamental
physics beyond the standard model. It will therefore be crucial to clarify
the property of this quark in various different aspects. In order to carry
out investigations for this purpose,
we have to analyze its decay processes by examining the final products, since
it turns into lighter particles right after being produced, even before the
hadronization, due to its huge mass.

In performing such studies, we have discovered a remarkable fact that the
angular distribution of the final charged-lepton $\ell^+$ ($\ell=e,\mu$) in
productions/decays of this quark depends only on possible nonstandard
``production'' interactions, in other words, it does not depend on any
nonstandard ``decay'' interactions at their leading order
\cite{Grzadkowski:1999iq}--\cite{Grzadkowski:2002gt},
see also \cite{Rindani:2000jg,Godbole:2006tq}. This is what we call
``the decoupling theorem in top productions/decays''. This theorem is valuable
in exploring new physics through analyzing possible
anomalous top-quark couplings, since we are thereby able to study its production
mechanism exclusively (i.e., without being affected by its decay interactions)
via the $\ell^+$ angular distribution. It is therefore meaningful to understand
this theorem from more than one viewpoint.

We have not found any problem in our proof of this theorem
\cite{Grzadkowski:1999iq}--\cite{Grzadkowski:2002gt}, but we have to admit that
we have not answered questions like ``Can you explain it in an intuitive or
visual way without using detailed calculations/formulas?'', which we received
many times after we published our papers. In order to compensate for this point,
I will see in this article to what extent we can understand it without relying
on any specific detailed calculations. Its original form is represented in terms
of the initial-state momentum in top productions as the reference axis. This,
however, makes visual arguments quite hard. Pointing out that we can study the
theorem through polarized top-quark decays, I aim here to present some clear
picture on how this theorem is born, which must be quite useful for other
heavy-quark phenomenology and also instructive.

\sec{Basic framework and strategy}

My strategy is to consider the theorem via decays of a polarized top quark, as
mentioned in the first section. Generally, extracting the decay part from whole
production/decay processes and treating that part independently is not justified,
but it is possible in our case because the narrow-width approximation is expected
to work well for the top-quark propagator. There, the top-quark spin direction is
completely decided by the production process alone. Hence, if we can show that
the angular distribution of the final particle around the top spin is not
dependent on any nonstandard couplings responsible for the decay, we can in fact
conclude that the angular distribution of the decay product of the top produced in
the process is not affected by these nonstandard couplings, which means the
theorem holds.

Let me show the base framework, which is the same as what we utilized in
\cite{Grzadkowski:1999iq}--\cite{Grzadkowski:2002gt}: Once a top quark is produced,
it decays immediately as $t \to b W^+$ in almost all cases. For describing those
processes, I use the most general $tbW$ coupling
\begin{eqnarray}
&&{\mit\Gamma}^{\mu}_{tbW} 
=-\frac{g}{\sqrt{2}}\bar{u}_b(\mib{p}_b,s_b)
\Bigl[\:\gamma^{\mu}(f^L_1 P_L+f^R_1 P_R)           \nonumber \\
&&\phantom{{\mit\Gamma}^{\mu}_{tbW} 
           =-\frac{g}{\sqrt{2}}\bar{u}_b(\mib{p}_b,s_b)\Bigl[\:\gamma^{\mu}}
-\frac{i\sigma^{\mu\nu}k_{\nu}}{M_W}
(f^L_2 P_L+f^R_2 P_R)\:\Bigr] u_t(\mib{p}_t,s_t),~~~~~~~~
\label{Gbt}
\end{eqnarray}
where $g$ is the $SU(2)$ coupling constant, $k$ is the $W$-boson momentum, $P_{L/R}
\equiv (1 \mp \gamma_5)/2$, $f_{1,2}^{L,R}$ are form factors (treated as constants)
with $f_1^L=1$ and $f_1^R=f_2^{L,R}=0$ in the standard model (SM), while I assume
that the $W$ boson decays into a charged-lepton $\ell^+$ and the
corresponding neutrino $\nu_{\ell}$ via the standard $V-A$ coupling. In the
following, I set the $z$ axis in the direction of the top-quark spin vector
$\mib{s}_t$, and take it as the angular-momentum quantization axis. I neglect all
the fermion masses except $m_t$, although the theorem still holds for
$m_b \neq 0$ \cite{Grzadkowski:2002gt,Godbole:2006tq}.

\begin{figure}[H]
\begin{center}
\begin{minipage}{14cm}
\begin{center}
\includegraphics[width=8.8cm]{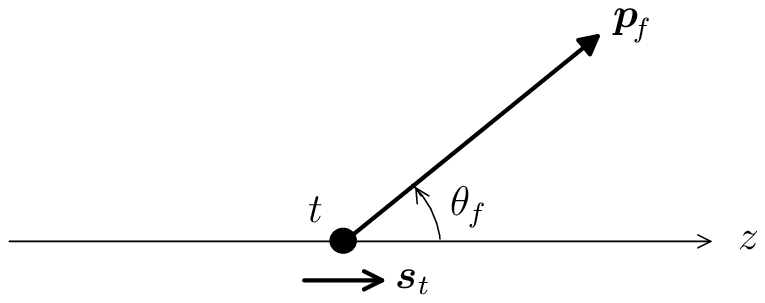}
\vspace*{0.35cm}
\caption{
Basic quantities describing the angular distribution of $f\,(=b,\ell^+,\nu_{\ell}$),
whose momentum is $\mib{p}_{\!f}$. The $z$ axis is set in the direction of the top-quark
spin vector $\mib{s}_{t}$, and it is taken as the angular-momentum quantization axis.
$\theta_f$ is the angle between $\mib{s}_{t}$ and ${\mib n}_f(\equiv {\mib p}_f/|{\mib p}_f|)$.
}\label{Basis}
\end{center}
\end{minipage}
\end{center}
\end{figure}

Now, I express the momentum of the final particle $f\,(=b,\ell^+,\nu_{\ell})$ as
${\mib p}_f$ and the unit vector of its direction as ${\mib n}_f$, i.e.
${\mib n}_f \equiv {\mib p}_f/|{\mib p}_f|$. Then, the angular distribution of $f$
becomes a function of ${\mib s}_t$ and ${\mib n}_f$, denoted as $F(\mib{s}_t,\:{\mib n}_f)$,
and this distribution must take the following form:\footnote{
    Note that $d{\mit\Gamma}$ depends on $\mib{s}_t$ at most linearly because
    this $\mib{s}_t$ appears there through $u(\mib{p}_t,s_t)\bar{u}(\mib{p}_t,s_t)
    \propto 1+\gamma_5 \sla{s}_{\!t}$.}\
\begin{equation}
d{\mit\Gamma}/d\cos\theta_f=F(\mib{s}_t,\:{\mib n}_f)
= C( 1 + P \mib{s}_t {\mib n}_f) 
= C( 1 + P \cos\theta_f ),
\label{Baseformula}
\end{equation}
where both $C$ and $P$ are constants\footnote{$P$ is indeed equivalent to the
    quantity known as ``the spin analyzing power", but it is a mere unknown
    parameter in our simple discussions here.}\ 
depending generally on $f_{1,2}^{L,R}$, and $\theta_f$ is the angle between
$\mib{s}_t$ and ${\mib n}_f$ as shown in Fig.\ref{Basis}. Since there is no
threshold coming from the momentum conservation in the processes we are considering,
$\cos\theta_f$ can vary from $-1$ to $+1$. Therefore, the full width becomes $2C$,
i.e., $C$ must be positive, which leads to the following constraint on $P$
\[
-1\ \leq\ P\ \leq\ +1,
\]
as $d{\mit\Gamma}/d\cos\theta_f$ must not be negative
for any $\cos\theta_f$.

\vskip 0.4cm

\vskip -0.3cm

Here, if our final goal is $d{\mit\Gamma}$ itself, we have to consider the anomalous-parameter
dependences of both $C$ and $P$. What interests us is however ``the $f$ distribution in whole
top-production plus decay processes'', where we need only $d{\mit\Gamma}$ normalized by
${\mit\Gamma}\,(=2C)$. We may therefore focus on $P$. There, if we can show that $P$ is free
from any anomalous $tbW$ couplings, it does mean that the decoupling theorem holds, because
they do not affect top-production processes as mentioned in the beginning. Generally, it will
be totally difficult to do this via our simple arguments alone, but I find there is one possibility.
That is to study if $P=\pm 1$ or not.
\\
{\bf $\mib{b}$-quark distribution}

\begin{figure}[H]
\begin{center}
\begin{minipage}{14cm}
\begin{center}
\includegraphics[width=8.0cm]{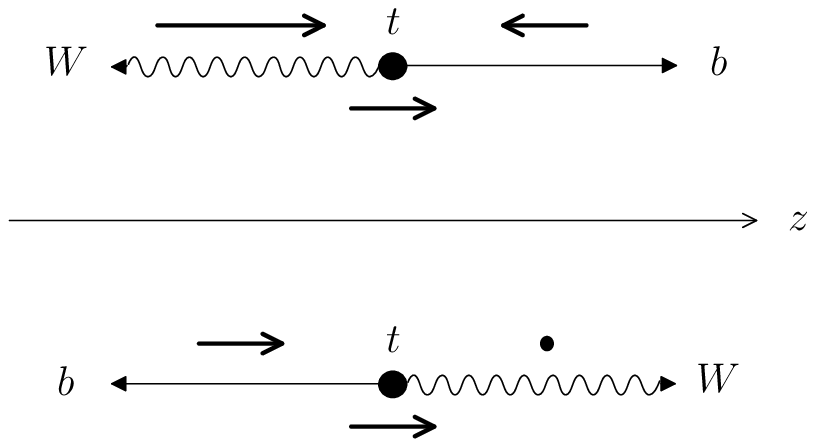}
\vspace*{0.35cm} 
\caption{
Allowed spin configurations for a left-handed $b$ quark to be emitted in the $\pm z$
directions in the top-quark rest frame, where the thick arrows express the $b$ and
$W$ spin vectors with nonvanishing $z$ components and the blob above the lower wavy
line expresses the $W$ spin vector whose $z$ component is zero.}\label{tbWFIG}
\end{center}
\end{minipage}
\end{center}
\end{figure}

Let me go over the $b$-quark angular distribution in our framework as a clear example.
Since my main concern here is in the leading nonstandard contributions coming from
the SM-coupling plus those which can interfere with it, I assume that the emitted $b$
is left-handed, to which $f_1^L$ and $f_2^R$ terms in eq.(\ref{Gbt}) contribute. As
we understand from Fig.\ref{tbWFIG}, the $b$ quark can be emitted both in the $+z$
direction ($\theta_b=0$) and $-z$ direction ($\theta_b=\pi$), depending on whether
$W^+$ is transverse or longitudinal. This means $P$ is neither $+1$ nor $-1$.
%
%
As mentioned, $P$ depends generally on the parameters of the decay interaction, i.e.,
$f_1^L$ and $f_2^R$ in this case, which shows that the decoupling theorem may not
hold for the $b$-quark distribution.\footnote{Through the simple arguments here alone,
    we cannot avoid the possibility that $P$ depends only on the SM coupling due to
    some reason. This is why I say ``$\cdots$ theorem may not hold $\cdots$''.}
This agrees with a conclusion of our preceding papers
\cite{Grzadkowski:1999iq}--\cite{Grzadkowski:2002gt} that the final $b$-quark does
not follow the theorem.

\vspace{0.6cm}

\vspace{-0.1cm}

\sec{Charged-lepton distribution}

Let us proceed to the charged-lepton distribution. It might seem possible to study it
the same way as in the previous section. It is indeed true that we can express the
$\ell^+$ angular distribution as
\begin{equation}
d{\mit\Gamma}/d\cos\theta_{\ell} = C( 1 + P \cos\theta_{\ell} )
\end{equation}
through arguments like those leading to eq.(\ref{Baseformula}). Then, if we thereby could
show that $P=+1$ (or $P=-1$), it means that the decoupling theorem holds. In this case,
however, it is never easy to develop similar analyses since we have to treat a
three-body final state. That is, when any of $\ell^+$, $b$ or $\nu_\ell$ is emitted in a
direction that is not parallel to the $z$ axis (the spin quantization axis), the state
of that particle becomes a superposition of $\ket{s_z=+1/2}$ and $\ket{s_z=-1/2}$ in
quite contrast to the classical mechanics, and we will no longer be able to carry out
clear discussions as for the $b$ distribution.

Therefore I would like to add another assumption: the parent top quark emits $W^+$ and $b$
parallel to its spin vector $\mib{s}_{t}$. Of course this does not always hold in the actual
$t$-decay process, however it is never that unreasonable as an approximation in order to
emphasize the characteristic feature of the process, since we can easily show according to
the spin conservation that these two particles are most likely emitted along this axis. 
Figure \ref{tbWFIG} must be helpful in understanding it in the case where $b$ is
left-handed, and I will also discuss this point in the next section.

\vspace{0.7cm}

\begin{figure}[H]
\begin{center}
\begin{minipage}{14cm}
\begin{center}
 {\includegraphics[width=14cm]{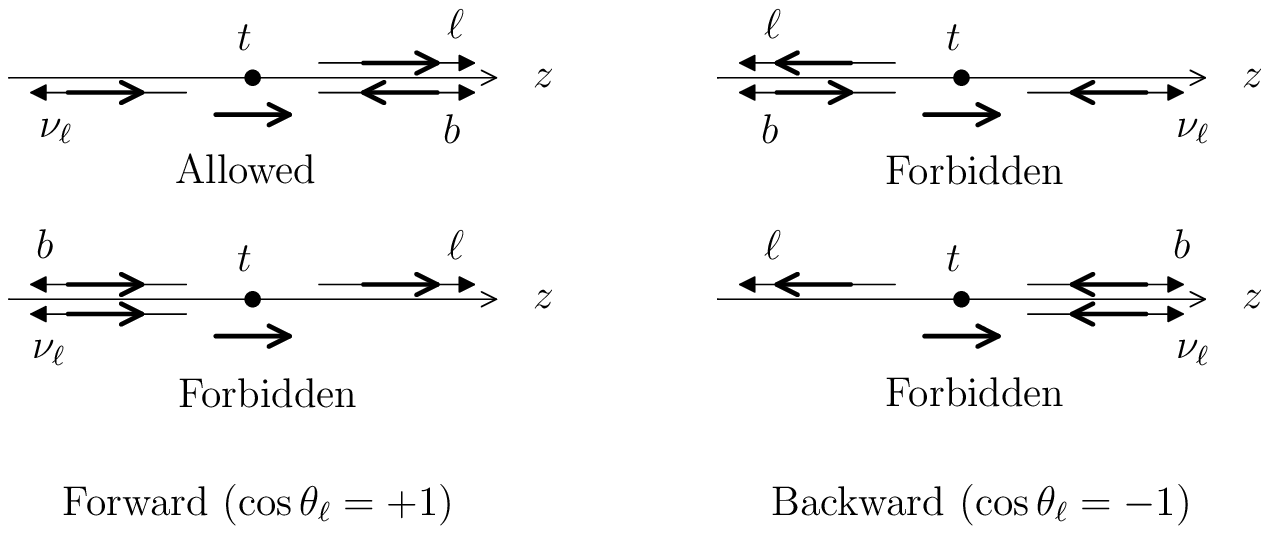}}
\vspace*{0.45cm}
\caption{
Allowed and forbidden spin configurations for the charged-lepton $\ell^+$ moving along
the $z$ axis in the top-quark rest frame, where the thick arrows express the spin vectors.
There might seem to exist other combinations in which $\ell$ and $\nu_{\ell}$ move together
in one direction, but such configurations are not realized through any $W$-boson boost in
$W \to \ell\nu_{\ell}$.
}\label{tblnFIG}
\end{center}
\end{minipage}
\end{center}
\end{figure}


Under this assumption, let us see whether $\ell^+$ could move in the $+z$/$-z$ directions.
Like the $b$ distribution, I first assume that the emitted $b$ is left-handed.
Since the final neutrino must also move parallel to the $z$ axis in this case, no orbital
angular momentum is involved.
Then, according to the linear-momentum conservation, we can draw four final-particle
configurations, and only one is allowed among them via the angular-momentum conservation
as shown in Fig.\ref{tblnFIG}.
This means that $\cos\theta_{\ell}=-1$ is not allowed and consequently $P$ must be $+1$:
\begin{equation}
d{\mit\Gamma}/d\cos\theta_{\ell} = C ( 1 + \cos\theta_{\ell} ).
\label{DTforell}
\end{equation}
It is noteworthy that the decoupling theorem holds beyond the first order in the anomalous
couplings in this situation.

Indeed the above result can be explicitly confirmed by the corresponding decay
formula. That is, the four diagrams in Fig.\ref{tblnFIG} are configurations
in which the energy of $\ell^+$ becomes the largest or the smallest, and the allowed
one is the case where it is the lowest possible one (and the $\nu_{\ell}$ energy
is the largest). As is given in the appendix, the differential width is expressed as
\begin{equation}
\frac{d{\mit\Gamma}}{d\varepsilon_{\ell}\, d\cos\theta_{\ell}}
\sim
f_+(\varepsilon_{\ell})(1+\cos\theta_{\ell})
       +f_-(\varepsilon_{\ell})(1-\cos\theta_{\ell})
\end{equation}
apart from an overall coefficient, where $\varepsilon_{\ell} \equiv E_\ell/m_t$,
and we have
\begin{equation}
f_+(\varepsilon_{\ell})
\propto
|\,r_{\ssW} f_1^L + f_2^R\,|^2\ \ \
{\rm and}\ \ \
f_-(\varepsilon_{\ell}) = 0
\end{equation}
for $\varepsilon_{\ell}=\varepsilon_{\ell}^{\rm min}(=r_{\ssW}^2/2 \simeq 0.11)$,
where $r_{\ssW} \equiv M_W/m_t$, in complete agreement with eq.(\ref{DTforell}).

Similarly, if $b$ is emitted via the $f_1^R$ or $f_2^L$ couplings and becomes right-handed,
the upper left configuration in Fig.\ref{tblnFIG} becomes ``forbidden" and the lower left
gets ``allowed",
which is the process where the $\ell^+$ energy becomes the largest. This result again
coincides with
\begin{equation}
f_+(\varepsilon_{\ell})
\propto
|\,r_{\ssW} f_1^R + f_2^L\,|^2\ \ \
{\rm and}\ \ \
f_-(\varepsilon_{\ell}) = 0
\end{equation}
for $\varepsilon_{\ell}=\varepsilon_{\ell}^{\rm max}(=0.5)$.

It will be interesting to note that our arguments using Fig.\ref{tblnFIG} still hold
even if $m_b$ is not neglected, since a nonvanishing $m_b$ could reverse the $b$-quark spin
in the right two graphs but those two remain to be ``forbidden". Correspondingly, we can
confirm
\begin{equation}
f_-(\varepsilon_{\ell}^{\rm max})=f_-(\varepsilon_{\ell}^{\rm min})=0
\end{equation}
for eq.(\ref{f-}), where $\varepsilon_{\ell}^{\rm max, min}$ are not the above ones but
those for $m_b \neq 0$ (see the appendix). This is consistent with
our proof that the decoupling theorem holds for $m_b \neq 0$ \cite{Grzadkowski:2002gt}
(see also \cite{Godbole:2006tq}).
This kind of simple arguments based on the spin-conservation is often seen in explaining
the form of the $\mu$ decay in the SM in the literature. Therefore, ours may seem a mere
extension of them, showing that such an observation also works for the most general $tbW$
couplings. Let us not forget, however, that understanding decay processes alone is not our
final purpose here.

Before closing this section, I check how adequate our approximation on the $W^+$ and $b$
directions which leads to the above arguments using the decay formula for $\varepsilon_{\ell}=
\varepsilon_{\ell}^{\rm max}$ and $\varepsilon_{\ell}=\varepsilon_{\ell}^{\rm min}$ is.
As shown, $f_-(\varepsilon_{\ell})$ vanishes for $\varepsilon_{\ell}^{\rm max,min}$
while is nonvanishing for a general $\varepsilon_{\ell}$, see eqs.(\ref{f0+},\ref{f0-}).
Therefore, if $f_-(\varepsilon_{\ell})$ increases sharply when $\varepsilon_{\ell}$
deviates from $\varepsilon_{\ell}^{\rm max}$ or $\varepsilon_{\ell}^{\rm min}$ even
a little,\ it means our arguments have made full use of the very special property of
$f_-(\varepsilon_{\ell})$, and consequently they get
\phantom{
--------------------------------------------------------------------------------------
}

\vfill 

\begin{figure}[H]
\begin{center}
\begin{minipage}{14cm}
\begin{center}
\includegraphics[width=11.5cm]{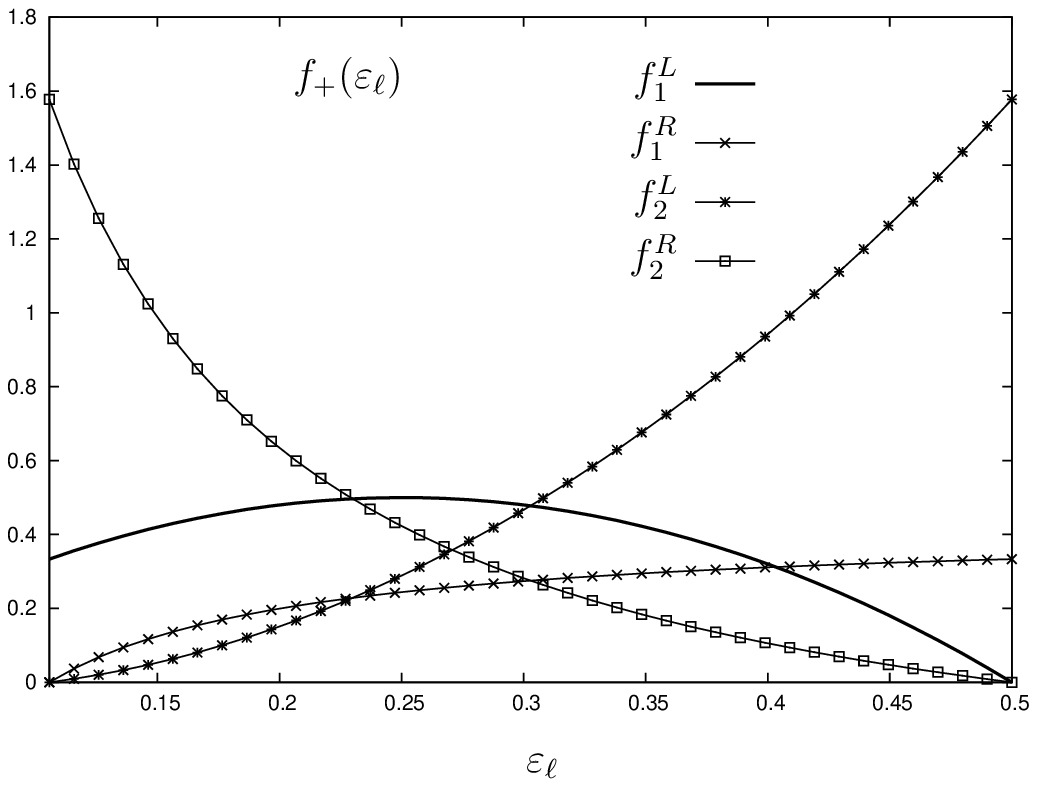}
\vspace*{0.0cm}
\caption{
Four curves show the contribution of the $f_{1,2}^{L,R}$ terms to $f_+(\varepsilon_{\ell})$.
For example, the curve named $f_1^L$ (solid curve) corresponds to $f_+(\varepsilon_{\ell})$
in which $f_1^L$ is set to be one and the others zero.
}\label{FIG-plus}
\end{center}
\end{minipage}
\end{center}
\end{figure}
\newpage
\begin{figure}[H]
\begin{center}
\begin{minipage}{14cm}
\begin{center}
\includegraphics[width=11.5cm]{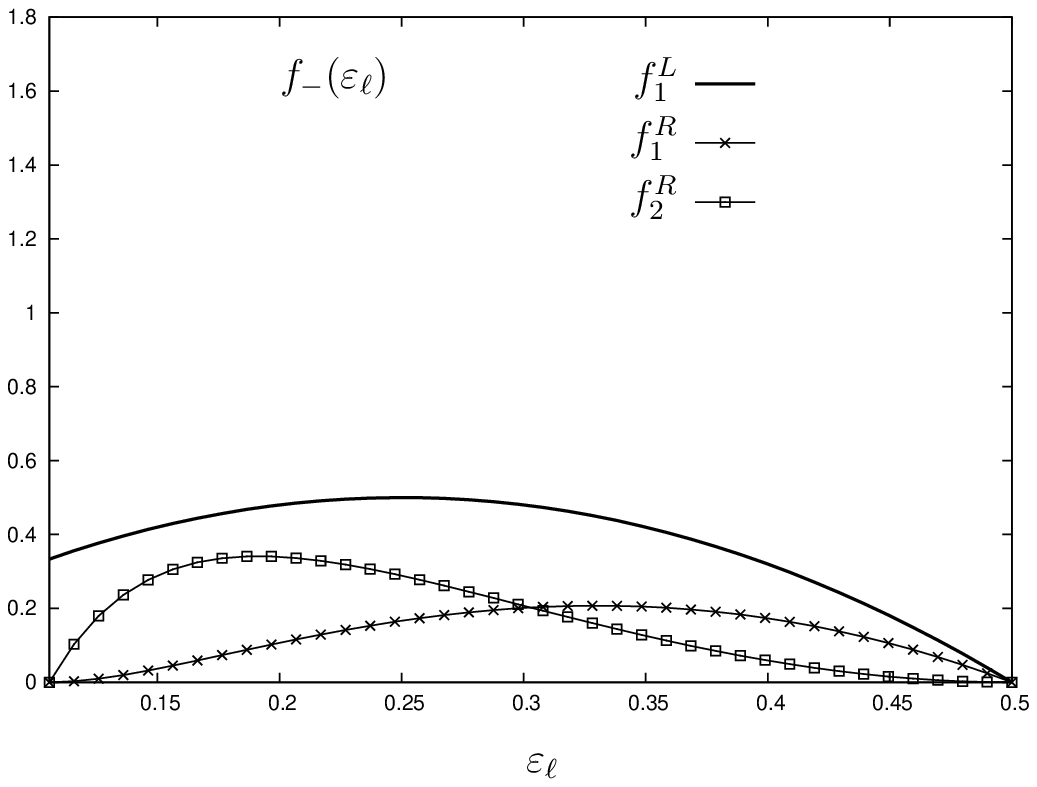}
\vspace*{0.0cm}
\caption{
Two curves named $f_{1,2}^R$ show their contribution to $f_-(\varepsilon_{\ell})$. The $f_1^L$
curve in $f_+(\varepsilon_{\ell})$ is also presented for comparison.
}\label{FIG-minus}
\end{center}
\end{minipage}
\end{center}
\end{figure}

\noindent
inappropriate for a general case.
Fortunately, however, it does not seem to be the case as seen in Figs.\ref{FIG-plus} and
\ref{FIG-minus}, where I have shown the $\varepsilon_{\ell}$ dependence of each term in
$f_{\pm}(\varepsilon_{\ell})$ using eqs.(\ref{f0+},\ref{f0-}): Figure \ref{FIG-plus} is
for $f_+$, and the curve named ``$f_1^L$" there, e.g., corresponds to $f_+$ in which $f_1^L$
is set to
be one and the others zero. Similarly Fig.\ref{FIG-minus} is for $f_-$, to which
only $f_{1,2}^R$ contribute but I have added the $f_1^L$ curve in $f_+$ for comparison.
These two figures tell us that the $f_{1,2}^R$ terms in $f_-$ remain small over the whole
range of $\varepsilon_{\ell}$.

\sec{More general cases}

What can we know on the $\ell^+$ distribution in a more general situation, in which the
momenta of the final $\ell^+$, $\nu_{\ell}$ and $b$ are not parallel to each
other? As mentioned in the preceding section, any state of those particles is
expressed as a superposition of its $s_z$ eigenstates $\ket{s_z=\pm 1/2}$ unless
its momentum is in the $+z$ or $-z$ direction. This quantum effect makes it totally difficult to
understand the theorem visually, but this never means that we can find nothing there.
In fact, we will be able to tell about a specific process
as ``favored/suppressed'' instead of ``allowed/forbidden''.

First, let us remember the left-handed $b$-quark distribution considered in section 2. It will
be then easy to understand that the theorem does not hold for the $W$ angular
distribution either since $W$ is emitted in the direction opposite to $b$ in
the top-quark rest frame. However the transverse $W$ with $helicity=-1$ (denoted
hereafter as $W_-$)
cannot be emitted in the $+z$ direction, i.e., $P=-1$ in eq.(\ref{Baseformula}),
and the longitudinal $W$
(denoted as $W_0$) cannot be emitted in the $-z$ direction, i.e., $P=+1$:
\begin{eqnarray}
&&d{\mit\Gamma}/d\cos\theta_{W_-} = C_- ( 1 - \cos\theta_{W_-} ),\label{Wminus} \\
&&d{\mit\Gamma}/d\cos\theta_{W_0} = C_0\; ( 1 + \cos\theta_{W_0}\, ).\label{W0}
\end{eqnarray}
This shows that the angular distributions of $W_-$ and $W_0$ are both free from
the anomalous decay-interaction couplings, i.e., they obey the decoupling theorem.

\begin{figure}[H]
\begin{center}
\begin{minipage}{14cm}
\begin{center}
 {\includegraphics[width=12cm]{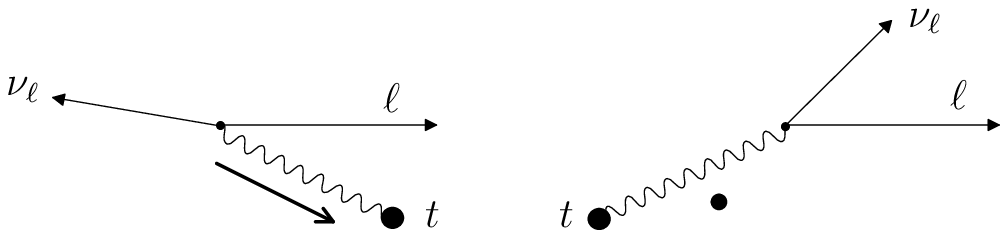}}
\vspace*{0.5cm}
\caption{
Favored spins and momenta configurations for a transverse $W^+$ (left side), a longitudinal
$W^+$ (right side), and $\ell^+$ in the top-quark rest frame. The $b$-quark line is not
shown for simplicity.
}\label{tblnFIG2}
\end{center}
\end{minipage}
\end{center}
\end{figure}

Now we know from the above equations that $W_-$ is likely to be emitted backward
while $W_0$ forward. Then we can study whether the final $\ell^+$ is likely/unlikely
to move in the $\pm z$ directions visually, noting that $\ell^+$ is most
likely to move in the $W$-boson spin direction in its rest frame. That is, $\nu_\ell$
tends to move in the same direction as the $W_-$ momentum and $\ell^+$ in the
opposite direction, while both of them from $W_0$
will move forward (but not parallel to each other) in the top rest frame. Figure \ref{tblnFIG2}
shows such configurations on the spins and momenta. Apparently, $\ell^+$ is likely(unlikely)
to be emitted forward(backward), indicating $P=+1$.

\vspace{0.4cm}

\begin{figure}[H]
\begin{center}
\begin{minipage}{14cm}
\begin{center}
 {\includegraphics[width=13.5cm]{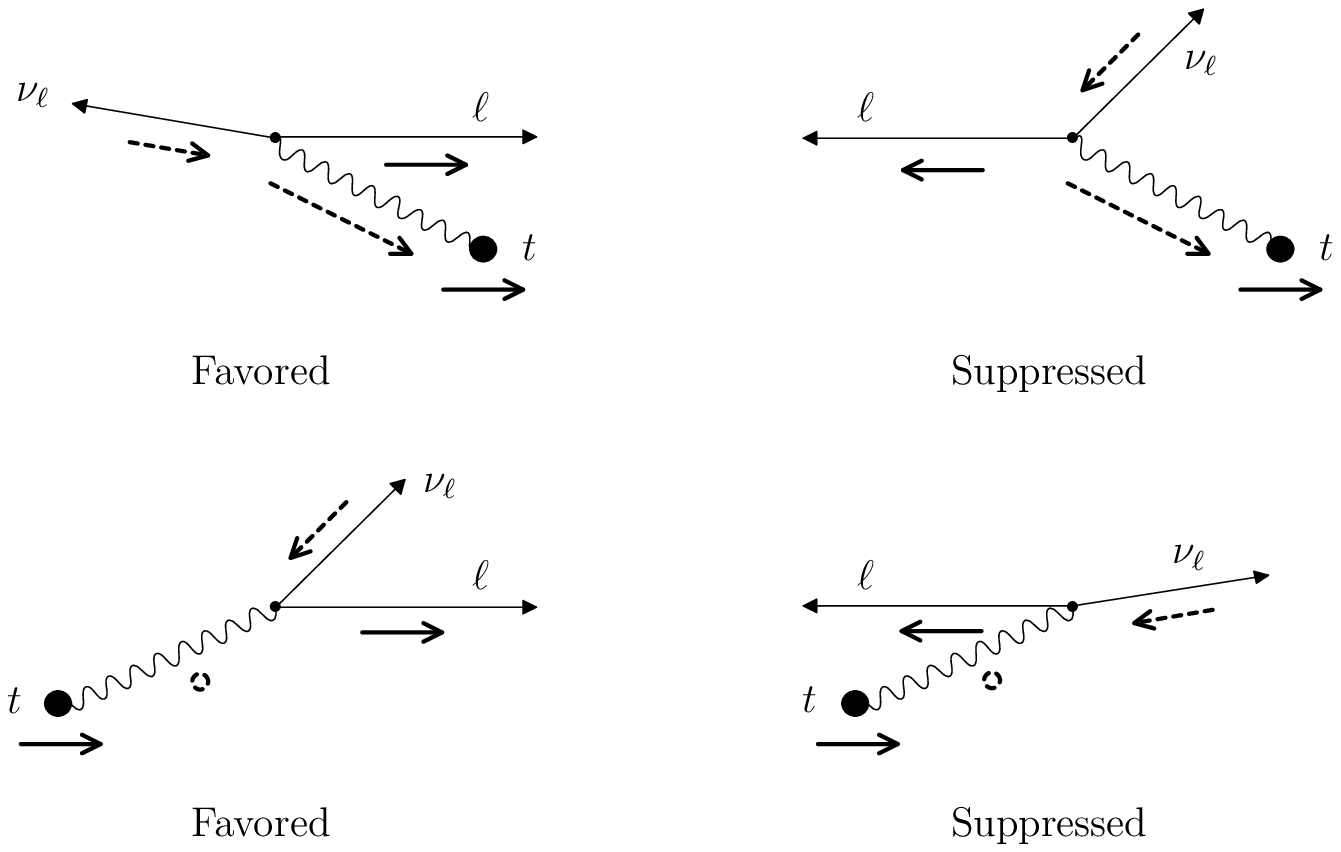}}
\caption{
Spins and momenta configurations for the charged-lepton $\ell^+$ emitted in the $+z$
direction (the left-side two diagrams) and $-z$ direction (the right-side two diagrams)
in the top-quark rest frame, where the thick solid arrows express the spin vectors in
the eigenstates, while those dotted-line arrows/circles mean that they are not in the
eigenstates. The $b$-quark line is not shown for simplicity.
}\label{tblnFIG3}
\end{center}
\end{minipage}
\end{center}
\end{figure}

We may say that those are analyses from a $W^+$-momentum viewpoint. We are also able to
examine the issue from a viewpoint of the spin ($z$ component) conservation. I show the
necessary spins/momenta configurations for $\ell^+$ emissions in the $\pm z$ directions
in Fig.\ref{tblnFIG3}, where the thick solid arrows express the spin vectors in the
eigenstates while those dotted-line arrows/circles mean that they are not in the eigenstates.
It will not be hard there to understand that the left/right two configurations are
favored/suppressed, taking
account of the spin conservation.\footnote{All the momenta in Fig.\ref{tblnFIG3} are on
    one common plane, so no orbital angular momentum (its $z$ component) is involved
    again.}\
In this way, we find it quite plausible for $\ell^+$ to move in the $+z$ direction but
not in the $-z$ direction.

\vspace{0.4cm}

That is, Figures \ref{tblnFIG2} and \ref{tblnFIG3} both support $P$ taking $+1$, which
indicates that the decoupling theorem still holds in this more general case. This also
seems to tell us that the same theorem is no longer valid for the $\nu_\ell$ angular
distribution since the lower left configuration is still valid while the upper right one
turns ``favored" if we exchange $\ell$ and $\nu_{\ell}$ in Fig.\ref{tblnFIG3}. We have not
studied the $\nu_\ell$ distribution in our preceding papers, but this agrees with the
results in \cite{AguilarSaavedra:2006fy}. Moreover, we can draw similar diagrams for
right-handed $b$ emissions. There, $W_+$($W$ with $helicity=+1$) and $W_0$ are likely
to be emitted forward and backward respectively, and neither $\ell^+$ emission in the
$-z$ direction (from $W_0$) nor that in the $+z$ direction (from $W_+$) is suppressed.

The above consideration can be applied to the standard-model term plus all the
standard-nonstandard interference terms and show qualitatively how the $\ell^+$
angular distribution gets to be proportional to $1+\cos\theta_\ell$, but do not
explain why the $|f_2^R|^2$ term does contribute both to $f_+$ and $f_-$ as in
eqs.(\ref{f0+},\ref{f0-}) in spite of
the fact that $b$ is also left-handed in the $f_2^R$ coupling. 
We need therefore some further analysis on this point.

First of all, note that the $f_2^{R}$ term in eq.(\ref{Gbt}) can be divided into two
different couplings, $\gamma^\mu (1 - \gamma_5)$ and $p_t^\mu (1 + \gamma_5)$, through
some $\gamma$-matrix algebra and Dirac equations of $t$ and $b$ (using $m_b=0$)
\cite{Mohammadi Najafabadi:2006um}. The
former structure is identical to $f_1^{L}$ term, while the latter gives the following
amplitude with the standard-model $\nu\ell W$ coupling:
\begin{equation}
{\cal M} \sim \bar{u}_b(\mib{p}_b,s_b) (1 + \gamma_5) u_t(\mib{p}_t,s_t) \cdot
\bar{u}_\nu(\mib{p}_\nu,s_\nu) \sla{p}_{\!t} (1-\gamma_5) v_\ell(\mib{p}_\ell,s_\ell)
\end{equation}
Although the lepton part includes $\sla{p}_{\!t}$, this is a kind of amplitude born
through a scalar-boson (denoted as $S$) production/decay, i.e., $\ell^+$ in this case is
a product of a leptonic $S$ decay.
Therefore, the final charged lepton is always produced via a polarized-vector ($W$) decay
in the $f_1^{L}$ coupling, while the one in the $f_2^{R}$ coupling receives contributions
both from a polarized-vector-boson decay and a scalar-boson decay. Since the latter is
not constrained by the same spin-polarization condition as $f_1^L$, then the pure $f_2^R$ term
could survive in $f_-$ through the ``scalar-exchange'' amplitude.
That is, the $|f_2^{R}|^2$ terms would break the theorem.

\sec{Summary}

The decoupling theorem in top-quark productions/decays is a valuable tool in
studying the property of this heavy quark in various aspects
\cite{Grzadkowski:1999iq}--\cite{Grzadkowski:2002gt},
\cite{Rindani:2000jg,Godbole:2006tq}. 
More specifically, this theorem is quite helpful in exploring new physics beyond the
standard model through analyzing possible anomalous top-quark couplings, since we
are thereby able to look into its production mechanism without being affected by its
decay interactions via the $\ell^+$ angular distribution. It is therefore quite meaningful
to grasp this theorem in many different ways, clarifying why such a theorem
could exist in simple and visual manners.

Focusing on semileptonic decays of a polarized top quark $t \to bW^+ \to b\ell^+ \nu_\ell$
instead of considering
full top production and decay processes, I have here tried to explain this theorem
without relying on any detailed calculations. I have shown first that the theorem
can be understood precisely as a result of the spin $z$ component conservation,
once we assume that $W^+$ and $b$ are emitted parallel to the top-quark spin vector
in $t \to bW^+$
as a reasonable approximation in order to emphasize the characteristic feature of
the process. Although I have treated the $b$ quark as a massless particle through
the main text for simplicity, those arguments are correct even when we do not neglect
its mass.

I then studied more general cases, in which the final three fermions
momenta are not parallel to each other. It is no longer possible to perform strict
arguments in those cases, but still I found that the decoupling theorem is quite
plausible from a $W^+$-momentum viewpoint and also from a spin-conservation
viewpoint.
There has been seen no contradiction between the present qualitative results and
previous quantitative ones.
I hope some visual picture on this theorem presented here is instructive and also
useful for other heavy-quark phenomenology.

\vskip 0.9cm

%
\secnon{Acknowledgments}
%
I would like to thank Bohdan Grzadkowski, with whom together we discovered the decoupling
theorem studied here, for kindly reading my notes, giving me many quite useful
comments on them from a variety of viewpoints, and also for suggesting that I publish
the notes. I am also grateful to Kazumasa Ohkuma for stimulating conversations
and for checking eqs.(\ref{t-wid})--(\ref{f-}).
This work was partly supported by the Grant-in-Aid for Scientific Research 
No.22540284 from the Japan Society for the Promotion of Science (by the end
of March, 2015).

\baselineskip=20pt plus 0.1pt minus 0.1pt

\vskip 0.8cm
\secnon{Appendix}

\vspace*{0.3cm}

I first present the angular and energy distribution of the final charged-lepton
($\ell^+$) coming from $t \to bW^+ \to b\ell^+ \nu_\ell$ in the top-quark rest
frame based on the interaction given in eq.(\ref{Gbt}) for the $tbW$ coupling
while the standard-model $V-A$ interaction for the $\nu\ell W$ coupling, keeping the $b$-quark
mass. Adopting the narrow-width approximation for the $W$-boson propagator,
the distribution is given as
\begin{equation}
\frac{d{\mit\Gamma}}{d\varepsilon_{\ell}\, d\cos\theta_{\ell}}=
\frac{g^4 m_t^2}{(32\pi)^2 r_{\ssW} {\mit\Gamma_W}}
\Bigl[\:f_+(\varepsilon_{\ell})(1+\cos\theta_{\ell})
       +f_-(\varepsilon_{\ell})(1-\cos\theta_{\ell})\:\Bigr],
\label{t-wid}
\end{equation}
where
\begin{eqnarray}
&&f_+(\varepsilon_{\ell}) =
4\, |f_1^L|^2
\varepsilon_{\ell} ( 1 -r_b^2 -2 \varepsilon_{\ell} ) 
+ |f_1^R|^2
r_{\ssW}^2 ( 2 - r_{\ssW}^2/\varepsilon_{\ell} ) \non\\
&&\phantom{f_+(\varepsilon_{\ell})}
-4\, |f_2^L|^2
\varepsilon_{\ell} ( r_{\ssW}^2 - 2 \varepsilon_{\ell} ) 
-|f_2^R|^2
r_{\ssW}^2 [\: 2 - ( 1 -r_b^2)/\varepsilon_{\ell} \:] \non\\
&&\phantom{f_+(\varepsilon_{\ell})}
-4\, {\rm Re}(f_1^L f_1^{R*}) r_b r_{\ssW}^2 
-8\, {\rm Re}(f_1^L f_2^{L*}) r_b r_{\ssW} \varepsilon_{\ell} \non\\
&&\phantom{f_+(\varepsilon_{\ell})}
+4\, {\rm Re}(f_1^L f_2^{R*}) r_{\ssW} ( 1 -r_b^2 -2 \varepsilon_{\ell} )
-4\, {\rm Re}(f_1^R f_2^{L*}) r_{\ssW} ( r_{\ssW}^2 - 2 \varepsilon_{\ell} ) \non\\
&&\phantom{f_+(\varepsilon_{\ell})}
-2\, {\rm Re}(f_1^R f_2^{R*}) r_b r_{\ssW}^3 /\varepsilon_{\ell}
-4\, {\rm Re}(f_2^L f_2^{R*}) r_b r_{\ssW}^2,
\label{f+} \\
%
%
&&f_-(\varepsilon_{\ell}) =
|f_1^R|^2 [\: -2 r_{\ssW}^2 (2-r_b^2 + r_{\ssW}^2)
+ 4(1-r_b^2 + 2r_{\ssW}^2) \varepsilon_{\ell} - 8 \varepsilon_{\ell}^2
+ r_{\ssW}^4 / \varepsilon_{\ell} \:] \non\\
&&\phantom{f_+(\varepsilon_{\ell})}
+ |f_2^R|^2 [\: 2 - 2(2-r_b^2)(r_b^2 - r_{\ssW}^2)
 -4 (2-2r_b^2 + r_{\ssW}^2 ) \varepsilon_{\ell} + 8 \varepsilon_{\ell}^2 \non\\
&&\phantom{f_+(\varepsilon_{\ell})}\ \ \ \ \ \ \ \
- r_{\ssW}^2 (1-r_b^2)/ \varepsilon_{\ell} \:] \non\\
&&\phantom{f_+(\varepsilon_{\ell})}
+ 2\, {\rm Re}(f_1^R f_2^{R*}) r_b r_{\ssW}
[\: - 2 ( 1 - r_b^2 + r_{\ssW}^2 ) + 4 \varepsilon_{\ell}
 + r_{\ssW}^2 /\varepsilon_{\ell} \:],
\label{f-}
\end{eqnarray}
$\theta_\ell$ being the angle between the $\ell^+$ momentum and the top-quark spin,
${\mit\Gamma_W}$ is the $W$-boson total decay width,
$r_b \equiv m_b/m_t$, $r_{\ssW} \equiv M_W/m_t$, $\varepsilon_{\ell} \equiv E_{\ell}/m_t$
with $E_{\ell}$ being the $\ell^+$ energy, and this ``normalized'' energy is restricted as
\[
(\varepsilon_{\ssW}-\kappa_{\ssW})/2 \leq \varepsilon_{\ell}
\leq (\varepsilon_{\ssW}+\kappa_{\ssW})/2
\]
with $\varepsilon_{\ssW}=(1-r_b^2+r_{\ssW}^2)/2$ and $\kappa_{\ssW}
= \sqrt{\varepsilon_{\ssW}^2 - r_{\ssW}^2}$\ \cite{Fujita-Hioki} (see also
\cite{Mohammadi Najafabadi:2006um}).
I have compared $d {\mit\Gamma}/d\cos\theta_{\ell}$ obtained by integrating eq.(\ref{t-wid})
on $\varepsilon_{\ell}$ with the corresponding formula in \cite{AguilarSaavedra:2006fy},
and confirmed that there is no discrepancy between them.

These equations show that all the leading contributions, i.e. those including $f_1^L$, are
only in $f_+(\varepsilon_{\ell})$ even when we keep $m_b$ finite.
That is, the angular distribution of the final charged lepton around the top-quark spin is
always proportional to $1+\cos\theta_{\ell}$ in the top-quark rest frame at the leading order
whatever form the $tbW$ coupling takes. This is equivalent to the decoupling theorem we found
in our previous papers \cite{Grzadkowski:1999iq}--\cite{Grzadkowski:2002gt} since any
top-decay interactions cannot affect the top-spin direction determined in production processes.

Let me next give the decay formula for a vanishing $b$-quark mass, which is directly related
to our arguments in the main text. Under this approximation, the forms of
$f_{\pm}(\varepsilon_{\ell})$ become simpler as
\begin{eqnarray}
&&f_+(\varepsilon_{\ell}) =
4\, |f_1^L|^2
\varepsilon_{\ell} ( 1 -2 \varepsilon_{\ell} ) 
+ |f_1^R|^2
r_{\ssW}^2 ( 2 - r_{\ssW}^2/\varepsilon_{\ell} ) \non\\
&&\phantom{f_+(\varepsilon_{\ell})}
-4\, |f_2^L|^2
\varepsilon_{\ell} ( r_{\ssW}^2 - 2 \varepsilon_{\ell} ) 
-|f_2^R|^2
r_{\ssW}^2 (2 - 1/\varepsilon_{\ell}) \non\\
&&\phantom{f_+(\varepsilon_{\ell})}
+4\, {\rm Re}(f_1^L f_2^{R*}) r_{\ssW} ( 1 -2 \varepsilon_{\ell} )
-4\, {\rm Re}(f_1^R f_2^{L*}) r_{\ssW} ( r_{\ssW}^2 - 2 \varepsilon_{\ell} ), 
\label{f0+}\\
%
%
&&f_-(\varepsilon_{\ell}) =
|f_1^R|^2 [\: -2 r_{\ssW}^2 (2+ r_{\ssW}^2)
+ 4(1+ 2r_{\ssW}^2) \varepsilon_{\ell} - 8 \varepsilon_{\ell}^2
+ r_{\ssW}^4 / \varepsilon_{\ell} \:] \non\\
&&\phantom{f_+(\varepsilon_{\ell})}
+ |f_2^R|^2 [\: 2 +4 r_{\ssW}^2
 -4 (2+ r_{\ssW}^2 ) \varepsilon_{\ell} + 8 \varepsilon_{\ell}^2
- r_{\ssW}^2 / \varepsilon_{\ell} \:],
\label{f0-}
\end{eqnarray}
and we have $\varepsilon_{\ssW}=(1+r_{\ssW}^2)/2$, $\kappa_{\ssW}=(1-r_{\ssW}^2)/2$,
leading to $\varepsilon_{\ell}^{\rm max}=1/2$ and
$\varepsilon_{\ell}^{\rm min}=r_{\ssW}^2/2$ with
\begin{eqnarray}
&&f_+(\varepsilon_{\ell}^{\rm max}) =
2 ( 1 - r_{\ssW}^2 ) |\,r_{\ssW} f_1^R + f_2^L\,|^2,\ \ \ \ \ \
f_-(\varepsilon_{\ell}^{\rm max}) = 0,
\label{f-max}
\\
&&f_+(\varepsilon_{\ell}^{\rm min}) =
2 ( 1 - r_{\ssW}^2 ) |\,r_{\ssW} f_1^L + f_2^R\,|^2,\ \ \ \ \ \
f_-(\varepsilon_{\ell}^{\rm min}) = 0,
\end{eqnarray}
which have been used in section 3.

Finally, there should be one comment on the narrow-width approximation for the
$W$-boson propagator, which we have adopted through our work as mentioned in the
beginning. This approximation is indeed quite helpful in calculations of the
$\ell^+$ distribution, but the resultant structure that all the leading terms in
the anomalous couplings are proportional to $1+\cos\theta_{\ell}$ is unchanged
even if we do not use it, as studied in \cite{Godbole:2006tq,Mohammadi Najafabadi:2006um}.

\vspace*{1.5cm}

\end{document}